\documentclass[article]{IEEEtran}
\usepackage[justification=centering]{caption}
\usepackage{booktabs}
\usepackage{makecell}
\usepackage{algorithm}
\usepackage{algorithmic}
\usepackage{amsmath}
\usepackage{amsthm}
\usepackage{amssymb}
\usepackage{caption}
\usepackage{graphicx, subfigure}
\usepackage{epstopdf}
\usepackage{algorithm}
\usepackage{algorithmic}
\usepackage{stfloats}
\usepackage{url}
\usepackage{subfigure}

\DeclareMathOperator*{\argmax}{argmax}
\begin{document}
\title{MSDF: A Deep Reinforcement Learning Framework for Service Function Chain Migration}
\author{\IEEEauthorblockN{Ruoyun Chen, Hancheng Lu, Yujiao Lu, Jinxue Liu\\
Department of Electrical Engineering and Information Science,\\
University of Science and Technology of China, Hefei, Anhui 230027 China\\
chenryun@mail.ustc.edu.cn, hclu@ustc.edu.cn, lyj66@mail.ustc.edu.cn, jxliu18@mail.ustc.edu.cn}\vspace{-2em}}
\maketitle
\begin{abstract}
Under dynamic traffic, service function chain (SFC) migration is considered as an effective way to improve resource utilization. However, the lack of future network information leads to non-optimal solutions, which motivates us to study reinforcement learning based SFC migration from a long-term perspective. In this paper, we formulate the SFC migration problem as a minimization problem with the objective of total network operation cost under constraints of users' quality of service. We firstly design a deep Q-network based algorithm to solve single SFC migration problem, which can adjust migration strategy online without knowing future information. Further, a novel multi-agent cooperative framework, called MSDF, is proposed to address the challenge of considering multiple SFC migration on the basis of single SFC migration. MSDF reduces the complexity thus accelerates the convergence speed, especially in large scale networks. Experimental results demonstrate that MSDF outperforms typical heuristic algorithms under various scenarios.
\end{abstract}
\begin{IEEEkeywords}
Service function chain, migration strategy, quality of service, multi-agents, deep reinforcement learning.
\end{IEEEkeywords}
\IEEEpeerreviewmaketitle
\section{Introduction}
Traffic flows in packet networks always require various services, which can be achieved by traversing kinds of middle-boxes. Traditionally, the network operator manipulates these middle-boxes with high cost due to specific hardware. Fortunately, network function virtualization (NFV) decouples network functions (NFs) from the underlying hardware so that various NFs can run as softwares on the commodity hardware to substitute the traditional middle-boxes \cite{mijumbi2016network}, which brings great flexibility to provide specific services thus significantly reduces the cost and complexity. Usually, a series of virtualized network functions (VNFs) are concatenated by virtual links in sequence to form the service function chain (SFC). Specifically, the corresponding VNF instances (VNFIs) hosted by virtual machines (VMs) will be deployed on servers in the data center network (DCN) firstly. Then the SFC is mapped to these instances and physical links to achieve the promised level of service. Nowadays, NFV is usually combined with software defined network (SDN) to provide easy traffic steering. However, the high dynamic of traffic makes it difficult to meet users' quality of service (QoS) with low cost all the time, which might reduce resource utilization in DCN.

Traditionally, there are several mechanisms to handle the traffic dynamic \cite{gouareb2019placement}, \cite{zhang2017optimizing} in NFV-enabled networks: \emph{i) Horizontal scaling}: adding or removing virtualized resource, e.g., VNF instance. Rahman et al. \cite{tang2018dynamic} propose VNF placement algorithms to scale in/out VNF instances which dynamically reacts to traffic changes. \emph{ii) Vertical scaling}: reconfiguring the size of virtualized resource allocated to containers or VMs. The authors in \cite{Efficient} formulate the dynamic resource scaling of VNF management as an integer linear programming (ILP) problem, and propose a greedy algorithm to solve it. \emph{iii) VNF migration}: migrating the related VMs in interrupted or alive manner. Cho et al. \cite{real} migrate the VNFs by going through all those migration strategies satisfying the modeled constraints to achieve low network latency. Quang et al. \cite{quang2018single} formulate the VNF forwarding graph placement problem as an ILP problem and consider VNF migration in dynamic networks. In fact, VNF migration is the most effective solution which can be combined with the other two scaling mechanisms to further enhance the resilience of VNF.

Nevertheless, VNF migration involves transferring the internal states of VMs, which causes QoS degradation and is costly especially in alive manner \cite{eramo2017approach}. To avoid QoS degradation caused by VM shutdown and reduce the network cost, SFC migration can be considered. Specifically, SFC migration implies manipulating the traffic forwarding by changing the mapping between VNFs in SFC and VNFIs in physical network. However, in DCN with dynamic traffic, obtaining the optimal migration strategy is difficult. As the time-varying traffic pattern is unknown, directly adopting instantaneous optimal migration strategy may lead to frequent migrations back and forth, which induces serious overhead. The optimal migration strategy should be considered from a long-term vision.

However, the long-term optimal migration strategy needs future network information. Fortunately, deep learning offers a tool to predict future network information accurately and easily with sufficient history data. Tang et al. \cite{tang2019virtual} propose a real-time VNF migration algorithm using the deep belief network to predict future resource requirements. But training the practicable neural network is extremely eager for large amounts of training data. Reinforcement learning (RL) \cite{sutton2018reinforcement} shows superiority when facing shortage of training data. Guided by appropriate reward, the RL agent can adjust its policy online and achieve the long-term goal finally. Li et al. \cite{li2019reinforcement} propose a RL based algorithm to solve the NP-hard VNF scheduling problem, while H. J. Ku et al. \cite{ku2017study} propose to migrate the whole SFC using RL to make better use of physical resource. However, both of them limit the migration action space in a small scale and solve it by tabular Q-learning, which is not practical in real networks with complicated states and great quantities of migration strategies.

In this paper, we solve the multiple SFC migration problem which considers joint migration of multiple SFCs in dynamic DCN while guaranteeing users' QoS. The high complexity in large scale networks motivates us to solve the problem using deep reinforcement learning (DRL). With the ability of extracting features, DRL can handle problems with high dimensional states thus overcome the shortcomings of tabular Q-learning. Furthermore, an cooperative multi-agent framework is proposed to address the challenge of huge action space. In detail, the contributions of this paper are listed as follows:

1) We formulate the SFC migration problem with time-varying traffic to minimize the total network cost in a long time span. We consider and analyze the constraints of users' customized QoS, in terms of end-to-end delay and packet loss. Moreover, both energy consumption and migration overhead are involved for the formulation of the optimization objective.

2) To solve the problem, we first explore the situation of single SFC migration. We design the deep Q-network (DQN) based subagent for single SFC migration, where the reward function of the subagent is designed as piecewise function based on the formulated optimization objective. In this way, the subagent can adapt to the dynamic traffic loads.

3) On the basis of subagents designed for single SFC migration, we propose a Monitor and Successive Decision Framework (MSDF) to handle the challenge of huge action space when consider joint migration of multiple SFCs. MSDF divides the huge action space into much more smaller ones of subagents for single SFC migration, thus the complexity is reduced and the convergence speed is consequently accelerated. In MSDF, the joint migration strategy is formed by concatenating actions of these subagents.

4) We compare the performance of MSDF with typical heuristic algorithms by experiments using real trace data. The experimental results show that MSDF outperforms existing algorithms under various scenarios.

The rest of this paper is organized as follows. Section II describes the system model and illustrates the problem formulation. Then Section III proposes the learning framework from single subagent to cooperative multi-agents in detail. Section IV reports the performance evaluation. Finally, the paper is concluded in Section V.

\section{System Model and Problem Formulation} \label{sec:model}
\subsection{Network model}
A time-varying DCN is considered in this paper, which is modeled by several main elements from the perspective of NFV:
\begin{itemize}
\item \textbf{Physical network}. It is modeled as an undirected graph $G^P=(N^P\!,L^P)$. A function node $i \in N^P$ is characterized by its resource capacity $C_i$. It can accommodate VMs on which the VNFIs are hosted. The physical link $(i,j)\in L^P$ between node $i, j$ is characterized by the propagation delay $D_{ij}$ on it.
\item \textbf{VNF set} $F$. The set $F\!=\!(N\!AT,Firewall,\cdots)$ includes all demanded types of VNF in DCN. Each type of VNF $v\!\in\! F$ is characterized by $(t_v^p, t_v^c, t_v^d, D_v)$ where the elements denote the coefficients of processing delay, resource allocation delay, deployment delay and deployment cost, respectively.
\item \textbf{SFC set} $S$. Each SFC $q\in S$ is denoted by a set of ordered virtual nodes. It is described by an undirected graph $G^V\!\!=\!\!(V\!,L^V)$ where node $v\!\!\in\!\! V$ is a logic VNF and $(k,l)\!\in\! L^V$ is a virtual link between logic VNF node $k,l$. Each SFC is characterized by its bandwidth and maximum end-to-end delay $(B_q, D_q)$.
\item \textbf{Flow set} $R$. Each flow $f\!\!\in\!\! R$, denoted by the pair of source node and destination node $(Src_f, Dst_f)$, is characterized by its bandwidth and maximum end-to-end delay $(B_f, D_f)$.
\end{itemize}

\subsection{Migration model}
Each SFC represents a tailored service for users that the network operator offers. It serves multiple flows that have the same request of service level but own different pairs of source node and destination node. With SFC migration, we change the mapping between logic VNFs and VMs on physical nodes, as well as the forwarding of relevant flows.

The detailed migration process is elaborated here: \emph{i): Preparation of the target node.} The centralized SDN controller informs the migration target node to perform resource reallocation on the related VM, according to the bandwidth of migrated SFC. If there is no requested type of VNFI, deploy one in advance. \emph{ii): Informing the original node.} The controller informs the original node to submit all the information of flows related to the migrated SFC, including session state information and data packets. \emph{iii): Information transferring.} The relevant information and data stored in the controller will be transferred to the target node, once the migration target node is ready. \emph{iv): Updating flow table.} The controller updates flow tables of related switches to forward flows according to the new mapping of SFC.

Note that if the source instance has no traffic to process after migration, the resource allocated to the related VM can be retrieved. If all the instances on the source node are idle, the node can run in a low power mode.

\subsection{Problem formulation}
\begin{table}[t]
\setlength{\abovecaptionskip}{0.1cm}
\caption{Variables of SFC migration problem}
\centering
\begin{tabular}{|p{0.05\textwidth}<{\centering}|p{0.37\textwidth}|}
\hline
Variables & \makecell[c]{Values}\\
\hline
\multicolumn{2}{|p{0.42\textwidth}<{\centering}|}{Time dependent variables}\\
\hline
$x_{i,v}^t$& Indicating the VNF of type $v\!\!\in\!\!F$ is deployed on function node $i\!\in\! N^P$ at time $t$ with value 1, 0 otherwise.\\
\hline
$y_{i,q,m}^t$& Indicating the VNF $m\!\!\in\!\! V_q$ of SFC $q\!\!\in\!\! S$ is mapped to node $i\!\!\in\!\! N^P$ at time $t$ with value 1, 0 otherwise.\\
\hline
$z_{ij,q,kl}^t$& Indicating the virtual link between neighboring $k_{th}$ VNF and $l_{th}$ VNF of SFC $q\in S$ ($(k,l)\in L^V$) is mapped to physical links between node $i$ and node $j$ ($(i,j)\!\!\in\!\! L^P$) at time $t$ with value 1, 0 otherwise.\\
\hline
$o_{i,v}^t$& Indicating the VM that hosts a VNFI of type $v\!\in\! F$ on node $i\!\in\! N^P$ has flows to process at time $t$ with value 1, 0 otherwise. From which we have $\sum_{\!v\in\! F}o_{i,v}^t\!\!=\!\!0$ indicating node $i$ is idle at time $t$.\\
\hline
\multicolumn{2}{|p{0.42\textwidth}<{\centering}|}{Time independent variables}\\
\hline
$h_{v,q,m}$& Indicating the VNF $m\in V_q$ of SFC $q\in S$ is type $v\in F$ with value 1, 0 otherwise.\\
\hline
$p_{q,f}$& Indicating service type of the flow $f\!\!\in\!\! R$ is translated into SFC $q\!\!\in\!\! S$ with value 1, 0 otherwise. From which we have $B_q\! =\! \sum_{\!f\in R}p_{q,f}B_f$ and $D_f\!=\!\sum_{\!q\in S}p_{q,f}D_q$.\\
\hline
\end{tabular}
\label{migvari}
\end{table}

We formulate the SFC migration problem which considers joint migration of multiple SFCs from a long-term perspective. The objective trades off between the operation cost and the revenue of guaranteeing users' QoS.

Above all, we define the long-term optimization time span as an operational cycle $T$, which is further divided into discrete time slots $t$ ($1\!\le\!t\!\le\!T,t\!\in\!\mathbb Z$). A migration decision will be made at the beginning of each time slot $t$. All the variables used in the SFC migration problem are binary, and they are summarized in Table \ref{migvari}.


To guarantee users' QoS, we consider two important metrics, i.e., maximum end-to-end delay $D_f$ and packet loss. The end-to-end delay includes the propagation delay on links and the fixed processing delay on VMs. There is a minimum processing resource request to make the processing speed catch with the arrival rate of data, otherwise packet loss will occur due to the lack of resource. These two metrics are represented by constraints, where the packet loss is guaranteed by the resource constraint.

\emph{1) End-to-end delay constraint}.
\begin{equation}\label{QoSCons}
\begin{aligned}
\sum_{q\in S}p_{q,f}\big(&\sum_{ij\in L^P}\!\sum_{kl\in L^V}z_{ij,q,kl}^tD_{ij}\!+\!D_{f^{\!S\!r\!c}0}\\&+D_{nf^{\!D\!s\!t}}+\!\!\sum_{m\in V}\sum_{v\in F}h_{v,q,m}t_v^p\big)\leq D_f,
\end{aligned}
\end{equation}
where $f^{\!S\!r\!c}$ and $f^{\!D\!s\!t}$ represent the source node and the destination node, respectively, while $0$ and $n$ represent nodes that are mapped by the first and the last VNF in $V_q$, respectively.

\emph{2) Resource constraint}. The sum of resource requested by all VNFIs on one node should not exceed the capacity of this node to avoid node congestion. We calculate the required computational capacity of VMs as \cite{eramo2016server}. Then we have the resource constraint as
\begin{equation}\label{resourCons}
\begin{aligned}
  \sum_{v\in F}C_{i,v}^t\leq C_i,
\end{aligned}
\end{equation}
where $C_{i,v}^t\!\!=\!\!\sum_{q\!\in\! S}\!B_q/\!{L\!e\!n}\!\sum_{m\in V_q}\!\!y_{i,q,m}^th_{v,q,m}t_v^p$ represents the requested resource of VM that hosts $V\!N\!F\!I_v$ on node $i$ at time $t$. $Len$ is the length of data packet. When the resource is insufficient, we allocate it following the max-min fair sharing algorithm and process packets according to the principle of first-come first-served. Then packet loss can be calculated as
\begin{equation}\label{pacloss}
PacLoss=\sum_{i\in N^P}\sum_{v\in F}\frac{C_{i,v}^t-R_{i,v}^t}{t_v^p}.
\end{equation}

With the above two constraints to guarantee users' QoS, we aim to minimize the total network cost in $T$, including energy consumption of the physical network and migration overhead.

\emph{1) Energy consumption}. The physical network consumes energy even when there is no traffic to process, which is called the basic resource consumption \cite{mijumbi2016network}. We separate it into two parts: the basic energy for running nodes and the basic resource consumption of VMs. To save the basic resource consumption, traffic processed on inefficient nodes should be aggregated, i.e., minimizing the number of running VMs and nodes. Thus we can express the energy consumption as:
\begin{equation}\label{enerCost}
\begin{aligned}
  E\!C\!O\!S\!T\!=\!\!\sum_{i\in N^{\!P}}\!\lambda_iInd(\!\sum_{v\in F}\!o_{i,v}^t)\!+\!\!\!\sum_{i\in N^{\!P}}\!\sum_{v\in F}\lambda_vo_{i,v}^t,
\end{aligned}
\end{equation}
where $Ind(x)$ is an indication function with value $1$ when $x\!\!\neq\!\!0$ and $0$ otherwise. The binary variable $o_{i,v}^t\!\!=\!\!1$ when $\sum_{q\!\in\! S}\sum_{m\!\in\! V_q}y_{i,q,m}^th_{v,q,m}\!>\!0$. $\lambda_i$ and $\lambda_v$ are the basic energy cost factors of node $i$ and VM that hosts the $V\!N\!F\!I_v$, respectively.

\emph{2) Migration overhead}. To keep the consistence, session state and data need to be stored and forwarded during the migration. Thus extra network resource is used during migration process, which is part of the migration overhead. The amount of these data is proportional to the product of the migration preparation time and the bandwidth of the SFC. By minimizing the following expression, frequent migrations can be avoided.
\begin{equation}\label{extrCost}
\begin{aligned}
  N\!C\!O\!S\!T\!\!=\!\!\!\sum_{q\in S}\!\sum_{m\in V_{\!q}}\!\sum_{i\in\!N^{\!P}}\!\!\!\Big(\!&\frac{B_q}2\!\!\left|y_{i,q,m}^t\!\!-\!\!y_{i,q,m}^{t\!-\!1}\right|\\&\!\!\!\times\!\!\!\sum_{v\in F}\!h_{v,q,m}\!\big(t_v^c\!\!+\!\!(1\!\!-\!\!x_{i,v}^t)t_v^d\big)\!\Big),
\end{aligned}
\end{equation}

Another part is reconfiguration cost including the cost of deploying new VNFIs and extra link bandwidth usage caused by link re-mapping, which can be expressed as follows.
\begin{equation}\label{configCost}
\begin{aligned}
  R\!C\!O\!S\!T\!\!=\!\!&\sum_{q\!\in\! S}\!\sum_{m\!\in\! V_{\!q}}\!\sum_{i\!\in\! N^{\!\!P}}\!\!\!\frac{\left|y_{\!i\!,q\!,m}^t\!\!-\!\!y_{\!i\!,q\!,m}^{t\!-\!1}\!\right|}2\!\!\sum_{v\!\in\! F}\!\!h_{\!v\!,q\!,m}\!(\!1\!\!-\!\!x_{\!i\!,v}^t\!)\!D_{\!v}\\
  &+\!\!\sum_{ij\!\in\! L^{\!P}}\!\sum_{kl\!\in\! L^{\!V}}\!\frac12\!\left|z_{ij,q,kl}^t-z_{ij,q,kl}^{t\!-\!1}\!\right|.
\end{aligned}
\end{equation}

The migration overhead is the sum of the above two parts, i.e., $M\!C\!O\!S\!T\!\!=\!\!\beta_nN\!C\!O\!S\!T\!+\!\beta_rR\!C\!O\!S\!T$. Where $\beta_n$ and $\beta_r$ are normalization factors.

Finally, the SFC migration problem is formulated in (\ref{P}) with the objective of minimizing the total network cost during $T$. In (\ref{cmb}), $\alpha_c,\beta_c$ are weights that balance the aforementioned two kinds of costs, satisfying $\alpha_c +\beta_c=1$. Constraint (\ref{c2}) indicates that each VNF should be mapped to one function node. (\ref{c3}) constrains the number of migrated VNFs for each SFC. (\ref{c4}) indicates that each type of VNF should be deployed on at least one function node. (\ref{c5}) ensures that each flow belongs to only one SFC. (\ref{c6}) guarantees the consistence of flow \cite{vizarreta2017qos}.

\begin{subequations}\label{P}
\setlength{\abovedisplayskip}{-2pt}
\setlength{\belowdisplayskip}{1pt}
\begin{align}
\label{cmb}
\mathbf P:min&\quad C\!O\!S\!T(\!T\!)\!=\!\!\sum_{t=1}^{T}\!\!\big(\alpha_{\!c\!} E\!C\!O\!S\!T(t)\!+\!\beta_{\!c\!} M\!C\!O\!S\!T(t)\big),\\
\label{c1}
s.t.&\quad (\ref{QoSCons}), (\ref{resourCons}),\\
\label{c2}
&\quad \sum_{i\in N^P}y_{i,q,m}^t=1,\\
\label{c3}
&\quad \sum_{m\in V_q}\sum_{i\in N^P}\frac12\left|y_{i,q,m}^t-y_{i,q,m}^{t-1}\right|\leq1,\\
\label{c4}
&\quad \sum_{i\in N^P}x_{i,v}^t\geq1,\\
\label{c5}
&\quad \sum_{q\in S}p_{q,f}=1,\\
\label{c6}
&\quad \sum_{ij\in L^P}\!z_{ij,q,kl}^t\!-\!\!\sum_{ji\in L^P}\!z_{ji,q,kl}^t=y_{i,q,k}^t\!-\!y_{i,q,l}^t.
\end{align}
\end{subequations}

It should be emphasized that the main decision variable of SFC migration is the VNF mapping variable: $y_{i,q,m},i\!\!\in\!\! N^{\!P}\!,q\!\!\in\!\! S,m\!\!\in\!\! V_q$, while other variables can be inferred based on previous network status.

\section{Monitor and Successive Decision Framework} \label{sec:MSDF}
The SFC migration problem is NP-hard as it is actually equivalent to the general case of the NP-hard bin packing problem \cite{hartmanis1982computers} by regarding the physical nodes as bins and VNFs as items. Moreover, the dynamic characteristic of the traffic makes it difficult to be solved by traditional heuristic methods. In this section, we propose a DRL based cooperative framework to approximate the optimal global solution in the time-varying DCN.

As the number of joint migration strategies is huge for a large scale network with multiple SFCs, we decompose the multiple SFC migration problem into multiple homogeneous single SFC migration problems. For each SFC, we design a DRL based subagent to make migration decisions. To form an effective joint migration strategy on the basis of these subagents, we design a cooperative framework to facilitate the minimization of total network cost when consider multiple SFC migration.


\subsection{Single SFC migration case}\label{SSFCM}
\begin{figure}[t]
\setlength{\belowcaptionskip}{-0.6cm}
  \centering
  \includegraphics[width=0.5\textwidth]{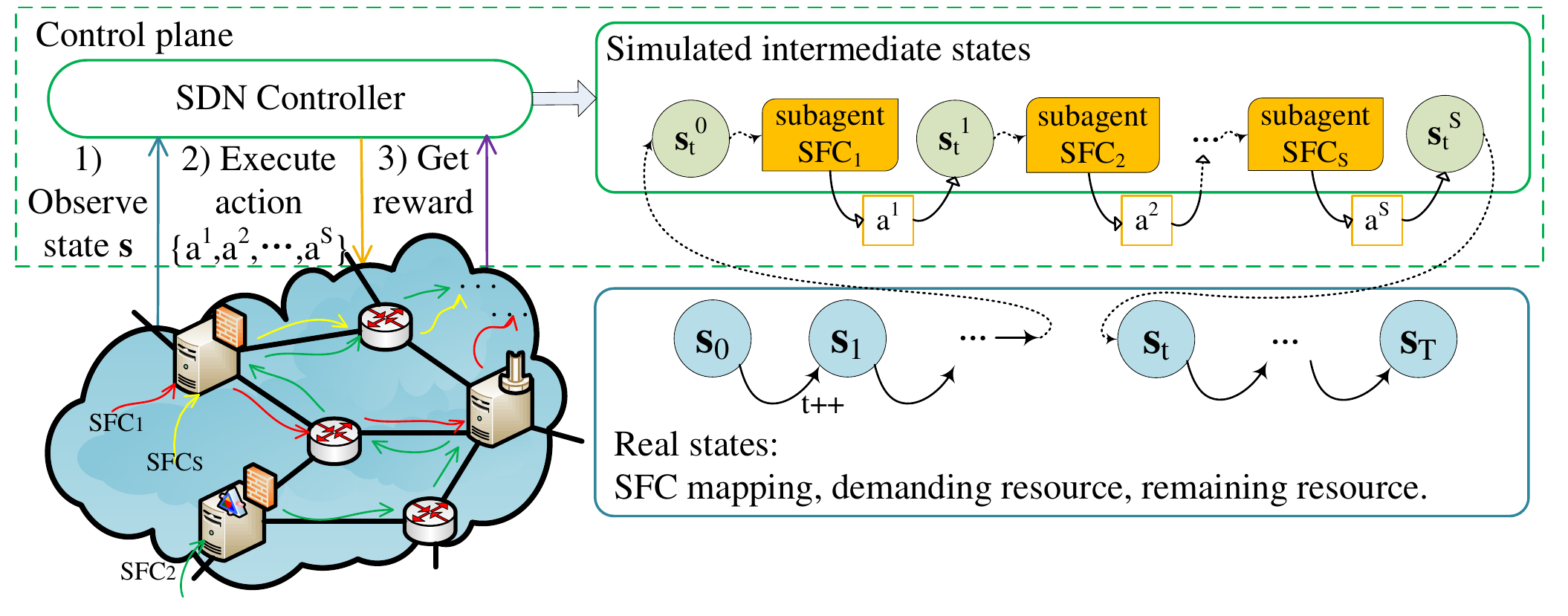}
  \caption{Overview of MSDF.}
  \label{multi}
\end{figure}

Since the current network status is only related to the current topology, traffic flow and the migration decisions, we consider the SFC migration problem as a Markov Decision Process (MDP). Adopting DQN \cite{dqn,dqnl}, we design the subagent for each SFC as follows, where we take the $q_{th}$ ($q\!\in\!S$) SFC as an example.


\textbf{\emph{State $\mathbf s$}}: The network state includes the resource requirements of VNFs in the $q_{th}$ SFC, the mapping of the $q_{th}$ SFC, and remaining resource. It can be expressed concisely as $\mathbf s_q^t \!=\!\big((b_1^t,\;b_2^t,\dots,\;b_{g_q}^t),(n_1^t,\;n_2^t,\dots,\;n_N^t)\big)$, where $b_i^t$ is the ratio of the demanding resource of the $i_{th}$ VNF in $S\!F\!C_{\!q}$ to the resource capacity of the mapping node, $n_i^t$ is the ratio of the remaining resource of the $i_{th}$ node to its resource capacity and $N$ is the number of available nodes in the DCN.

\textbf{\emph{Action $\mathbf a$}}: The action can be described as selecting one VNF from the VNF set of each SFC and deciding whether to migrate it and where to migrate it to, which forms a finite discrete action space.

\textbf{\emph{Reward $\mathbf r$}}: Considering the operation cost and users' QoS, the reward function is designed as the sum of (\ref{cmb}) and a punishment related to the constraints (\ref{c1}). As each subagent contributes to the migration overhead to varying degrees, we separately calculate the migration overhead of each subagent, and the global energy consumption is used to promote cooperation among subagents. Furthermore, in order to trades off among different optimization targets under different load scenarios, we design the reward as a piecewise function segmented by the degree of loads. Thus the reward function of the $q_{th}$ subagent can be expressed as (\ref{eqn:6}), where the $M\!C\!O\!S\!T_{\!q}(t)$ represents the migration overhead of the $q_{th}$ subagent.
\begin{equation}\label{eqn:6}
\begin{aligned}
\mathbf r_{\!q}(t)\!=\!\left\{\begin{array}{l}\!\!\!\!\--\big(\!\alpha_{\!c}E\!C\!O\!S\!T(t)\!+\!\beta_{\!c}M\!C\!O\!S\!T_{\!q}(t)\\
+\gamma_{\!c}P(t)\!\big),0\!\!<\!\!P(t)\!\!<\!\!\rho\\
\!\!\!\!-\big(\!\alpha_cE\!C\!O\!S\!T_{\!m\!a\!x}\!+\!\beta_{\!c}M\!C\!O\!S\!T_{\!q}(t)\\
+\gamma_{\!c}P(t)\!\big),P(t)\!\!\geq\!\!\rho\end{array}\right.,
\end{aligned}
\end{equation}
where $P(t)\!\!=\!\!f(\frac{real\!-\!max}{max})\!\!+\!\!f(PacLoss)$. $f(x)\!\!=\!\!max(0,\!x)$ is a unit ramp function. $real$ and $max$ are shorthand of real and maximum end-to-end delay. $\rho$ is a threshold of overload, which can be set according to the percentage of loads on nodes.

\begin{algorithm}[htbp]
\small
\caption{Training process of the $q_{th}$ subagent.}
\label{train-dqn}
\begin{algorithmic}[1]
\STATE Initialize $count=0$, initialize the constant $C$ to denote the period of updating the target network;
\WHILE{not converge}
\STATE Observe the state $s_t$ of simulated network.
\STATE Take action $a_t\!=\!\argmax_a\!Q_q(s_t,a)$ with exploitation probability $\epsilon$.
\STATE Observe the next state $s_{t+1}$ and reward $r_t$. Store the transition sample $(s_t,a_t,r_t,s_{t+1})$ in the memory;
\IF{$number\;of\;samples>batch\;size$}
\STATE Optimize the parameters $\theta^{Q_q}$ by using back propagation of the loss function \cite{dqn};
\STATE $count++$;
\IF{$count==C$}
\STATE {Update parameters of the target Q-network:\\
$\theta^{Q^{target}_q} \leftarrow \tau\theta^{Q_q}+(1-\tau)\theta^{Q^{target}_q}$}
\ENDIF
\ENDIF
\ENDWHILE
\end{algorithmic}
\end{algorithm}

\begin{algorithm}[htbp]
\small
\caption{Executing SFC migration using MSDF.}
\label{decision-dqn}
\begin{algorithmic}[1]
\STATE {Sort SFCs in descending order of $P_{q,t}^{S\!F\!C}$;}
\STATE {Initialize parameters of neural networks for all $S$ subagents: $Q(s,a;\theta^{Q_q})$ for the $q_{th},\;q\in S$ and $\theta^{Q^{target}_q}\!\leftarrow\! \theta^{Q_q}$;}
\WHILE{reward variance exceed threshold}
\FOR{$t=0$; $t<T$; $t++$ }
\STATE Observe the real network and take a snapshot of current network to create the simulated network;
\FOR{$q=0$; $q<S$; $q++$ }
\STATE Train the neural networks of $q_{th}$ subagent as Alg. \ref{train-dqn}. Apply the decision to the simulated network;
\ENDFOR
\STATE Apply joint decisions of all $S$ subagents to the real network;
\STATE Observe the next state $\mathbf s_{t+1}$ and reward $\mathbf r_t$;
\STATE $\mathbf s_t\!\leftarrow\!\mathbf s_{t+1}$;
\ENDFOR
\ENDWHILE
\end{algorithmic}
\end{algorithm}

DQN uses state-action value function $Q(s,a)$ to estimate the delayed instant return $r$, which evaluates the quality of action $a$ under state $s$. The optimal Q-function $Q^\ast(s,a)\!\!=\!\!max_{\mathrm\pi}E\lbrack R_t\vert s_t\!\!=\!\!s,a_t\!\!=\!\!a,\pi\rbrack$ estimates the maximum expected return from state $s$ with executing action $a$ thereafter following policy $\pi$, where $R_t\!\!\!=\!\!\!\sum_{t'\!=t}^T\!\gamma^{t'\!-t}r_{t'}$. To stabilize the learning, we adopt \emph{experience replay} \cite{dqn} which removes correlations between transition samples, and \emph{target network} \cite{dqnl} which keeps the consistency of targets when updating parameters. The training process is described as Algorithm. \ref{train-dqn}.




\subsection{Multiple SFC migration case}

\begin{figure*}[htbp]
\setlength{\belowcaptionskip}{-0.5cm}
\centering
\subfigure[one-agent v.s. multi-agent.]{
\begin{minipage}[t]{0.225\textwidth}
\centering
\includegraphics[width=1\textwidth]{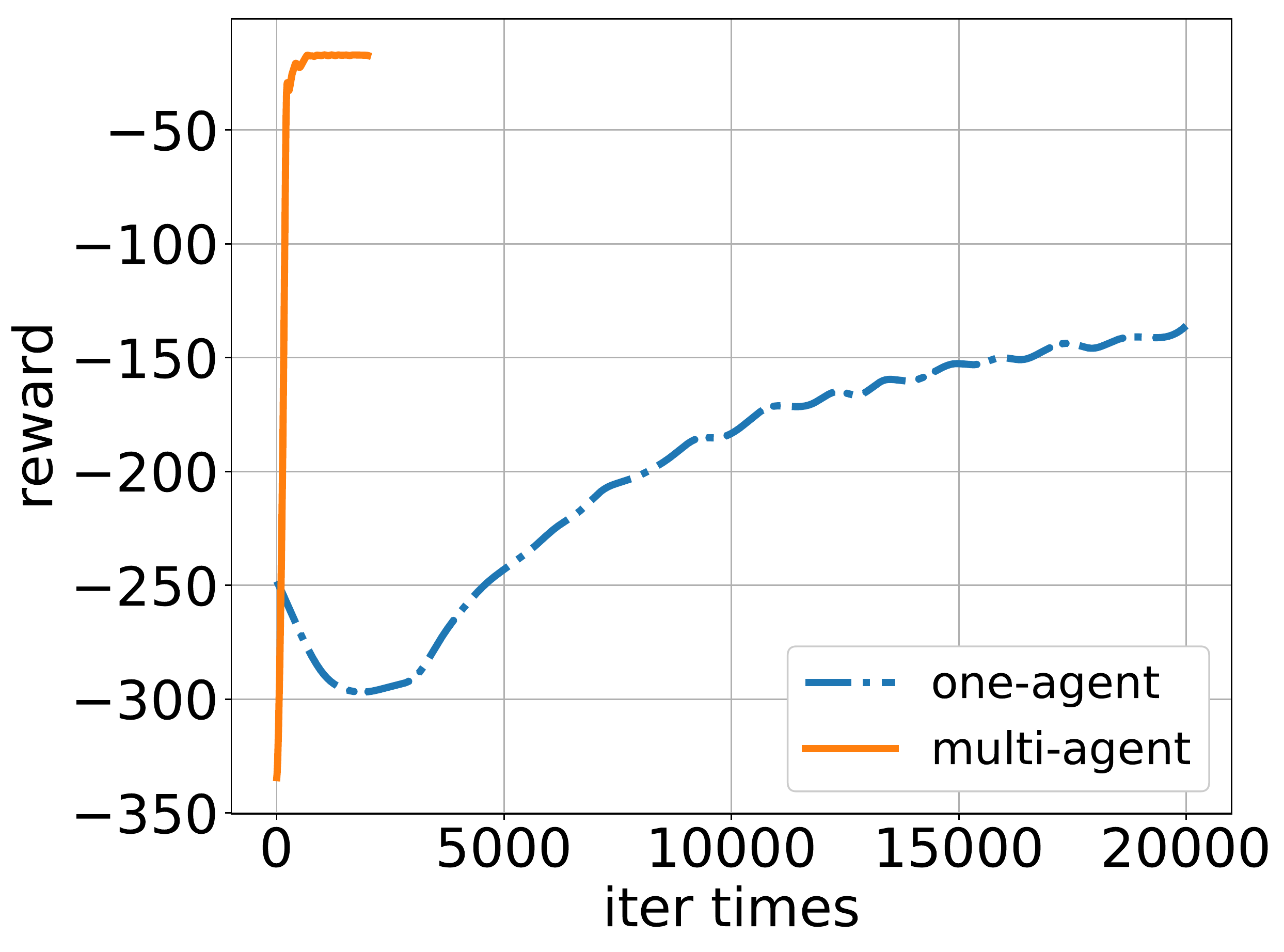}
\label{conv12m}
\end{minipage}
}
\subfigure[Number of SFCs: different number of cooperative subagents.]{
\begin{minipage}[t]{0.225\textwidth}
\centering
\includegraphics[width=1\textwidth]{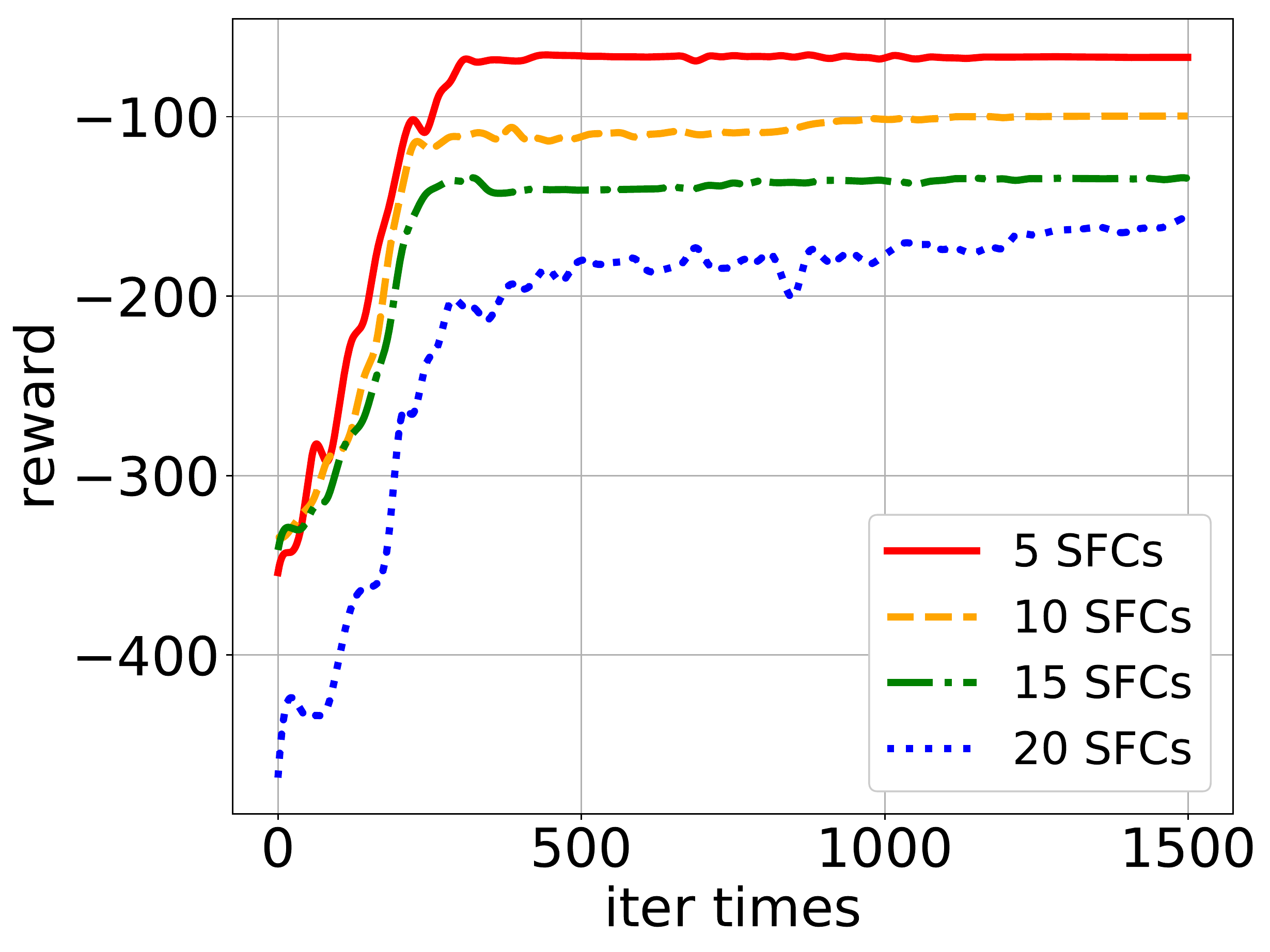}
\label{convsfcnum}
\end{minipage}
}
\subfigure[Number of VNFs: different size of action space.]{
\begin{minipage}[t]{0.225\textwidth}
\centering
\includegraphics[width=1.04\textwidth]{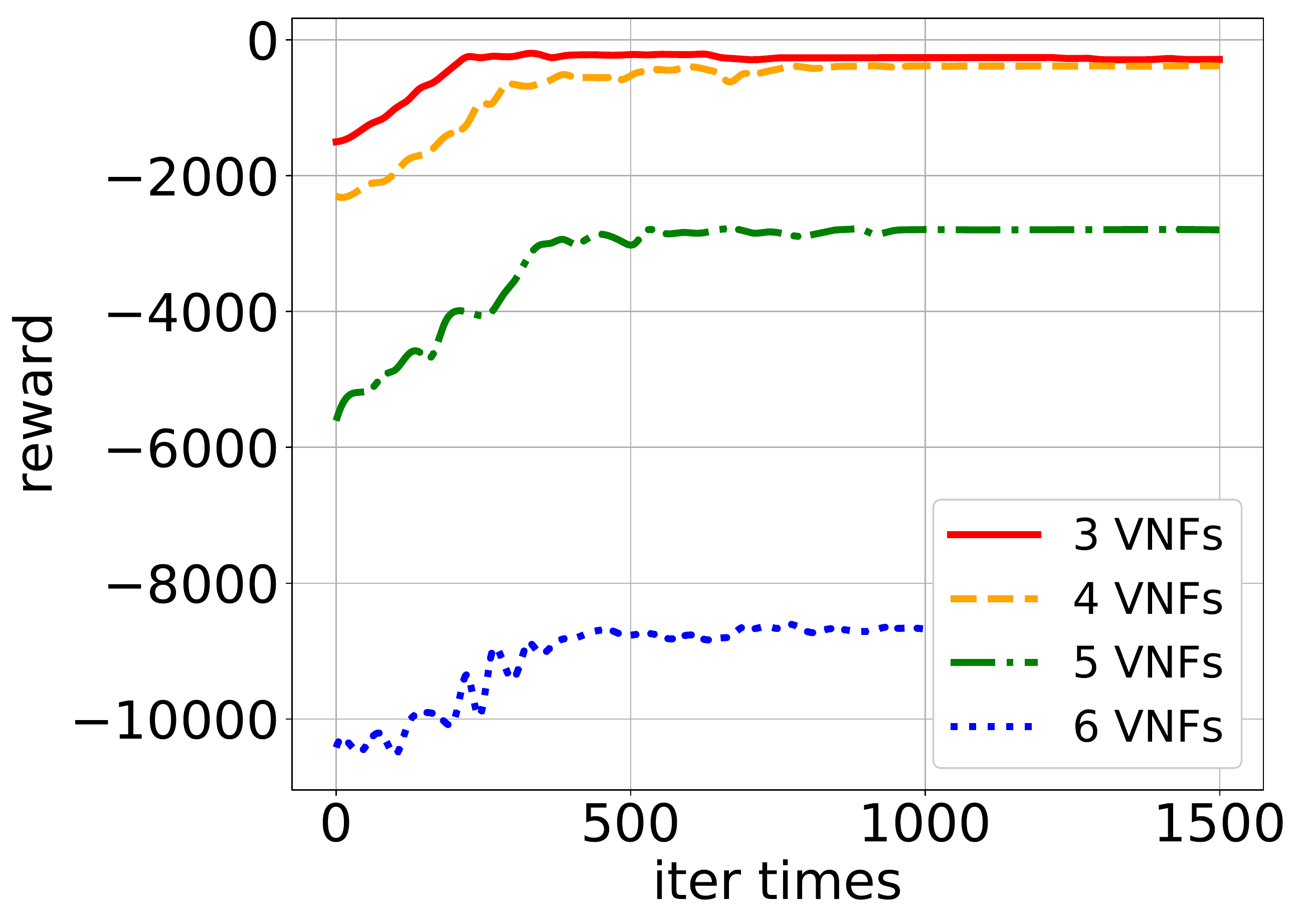}
\label{convsfclen}
\end{minipage}
}
\subfigure[Number of flows contained in each SFC: different targets.]{
\begin{minipage}[t]{0.225\textwidth}
\centering
\includegraphics[width=1.01\textwidth]{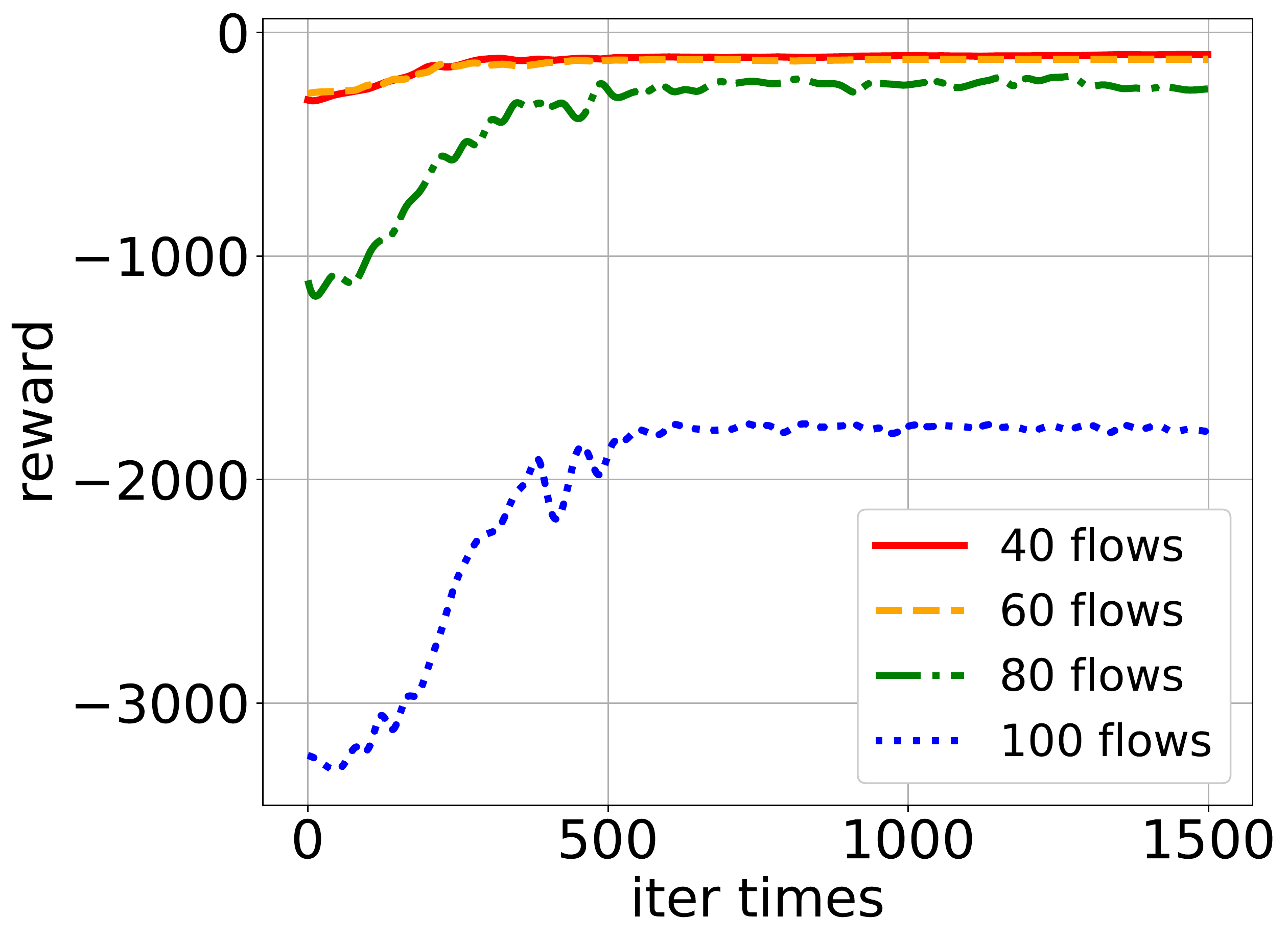}
\label{convflownum}
\end{minipage}
}
\centering
\caption{The convergence performance under different factors.}
\label{conver}
\end{figure*}

\begin{figure*}[htbp]
\centering
\subfigure[Total cost.]{
\begin{minipage}[t]{0.225\textwidth}
\centering
\includegraphics[width=0.97\textwidth]{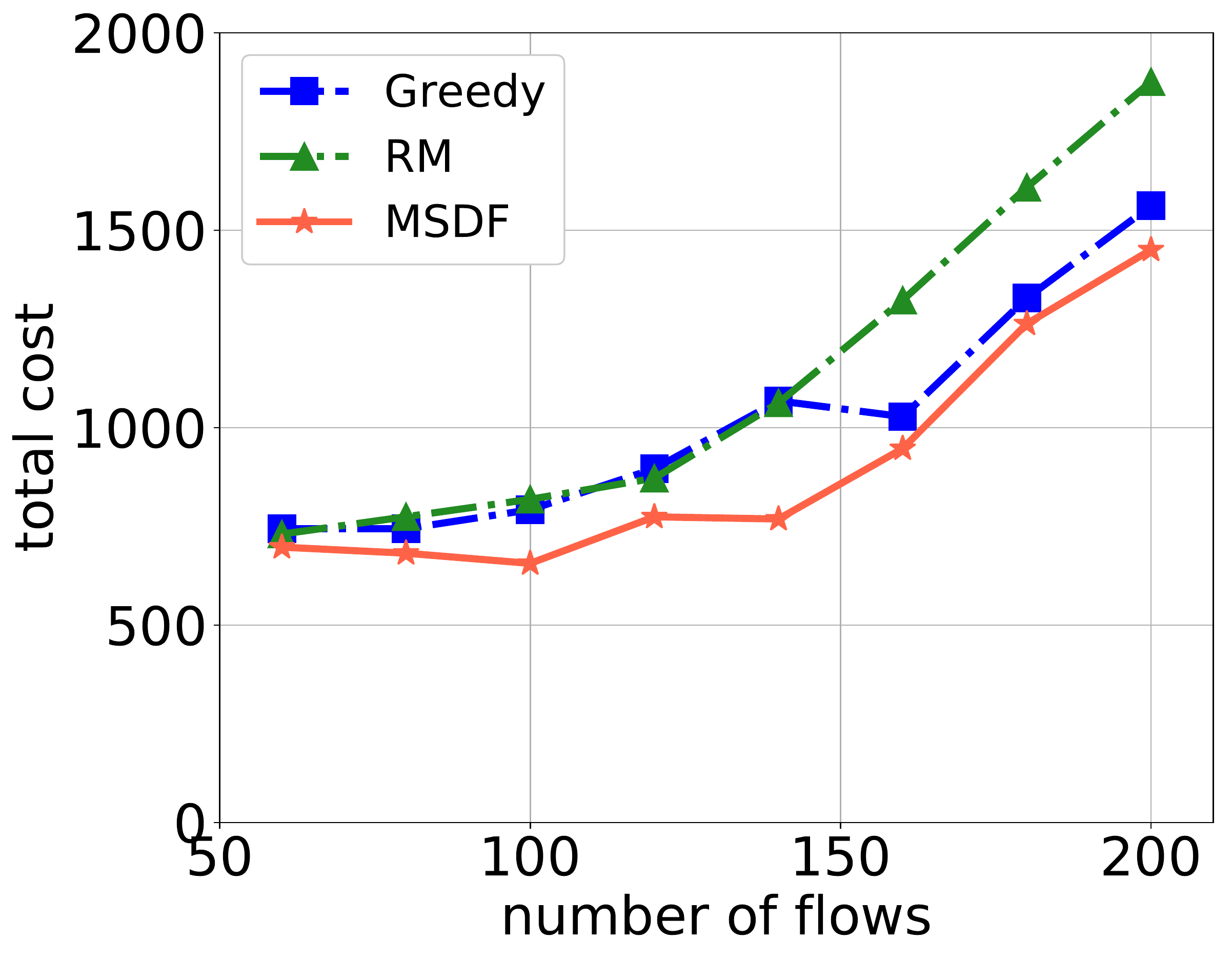}
\label{uu-f-t}
\end{minipage}
}
\subfigure[Number of migrations.]{
\begin{minipage}[t]{0.225\textwidth}
\centering
\includegraphics[width=0.93\textwidth]{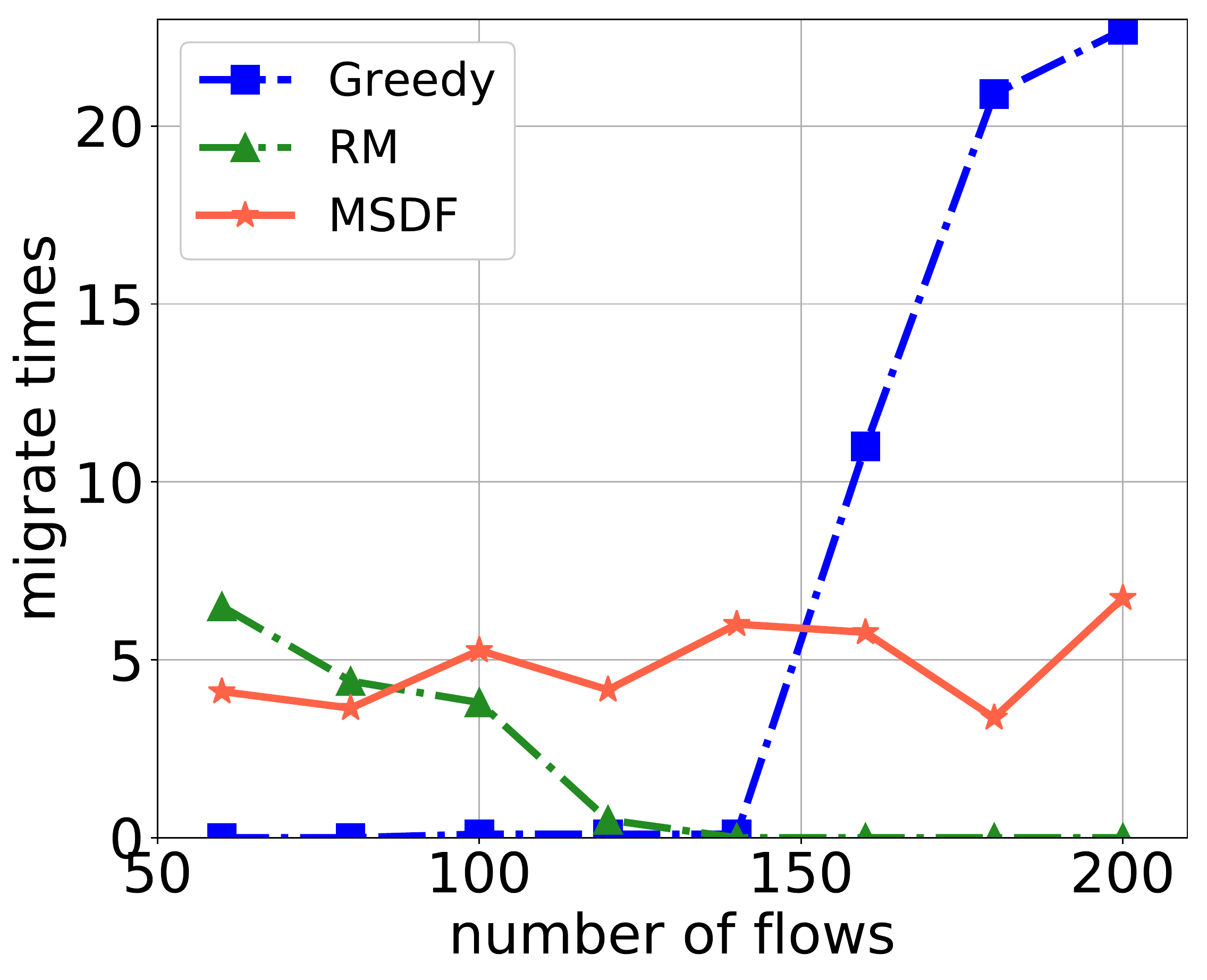}
\label{uu-f-m}
\end{minipage}
}
\subfigure[The degree of overload on nodes.]{
\begin{minipage}[t]{0.225\textwidth}
\centering
\includegraphics[width=0.94\textwidth]{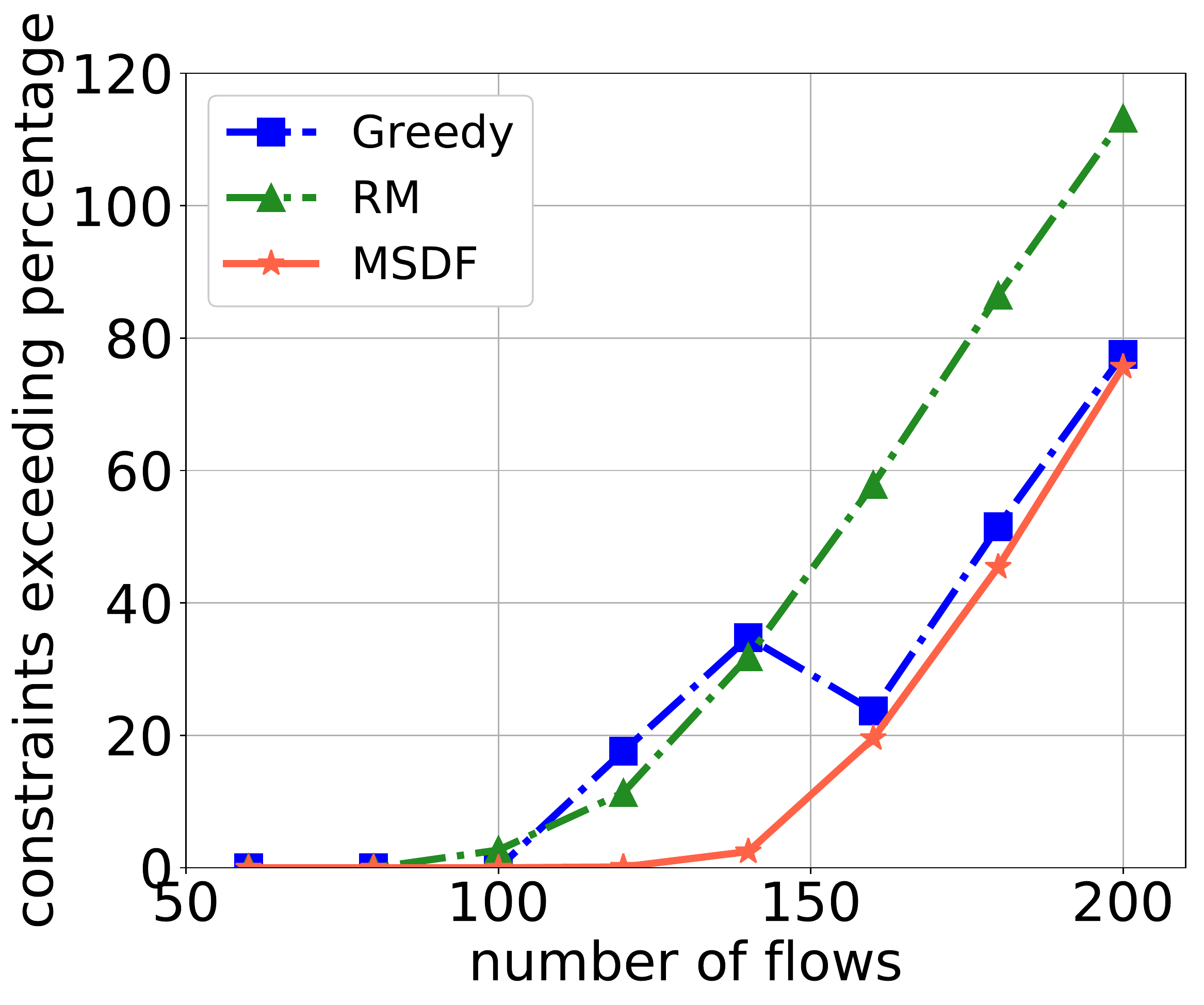}
\label{uu-f-o}
\end{minipage}
}
\subfigure[Basic energy cost.]{
\begin{minipage}[t]{0.225\textwidth}
\centering
\includegraphics[width=0.95\textwidth]{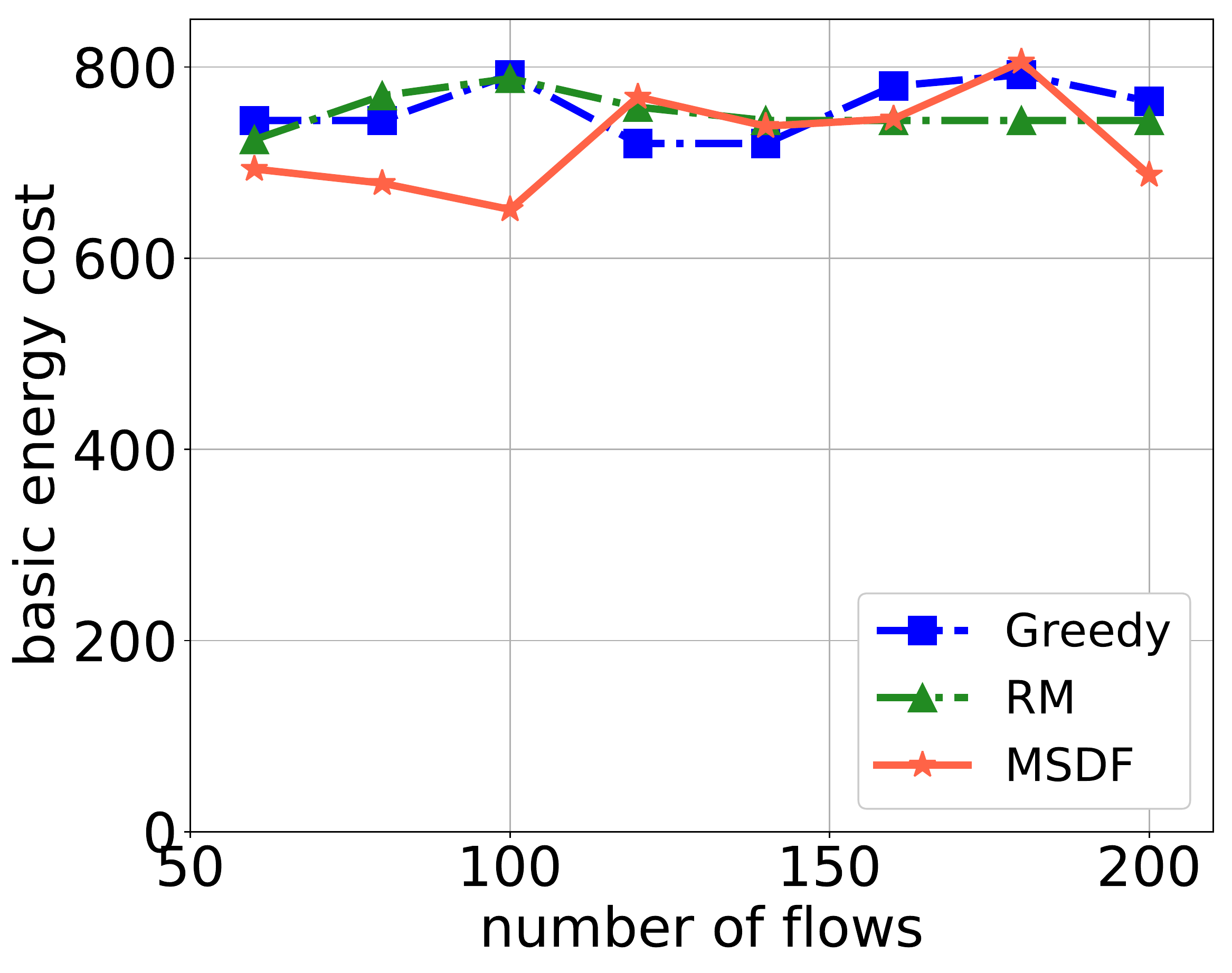}
\label{uu-f-e}
\end{minipage}
}
\centering
\caption{Performance changing along number of flows in each SFC.}
\label{uu-f}
\end{figure*}

\begin{figure*}[htbp]
\setlength{\belowcaptionskip}{-0.6cm}
\centering
\subfigure[Total cost.]{
\begin{minipage}[t]{0.225\textwidth}
\centering
\includegraphics[width=0.95\textwidth]{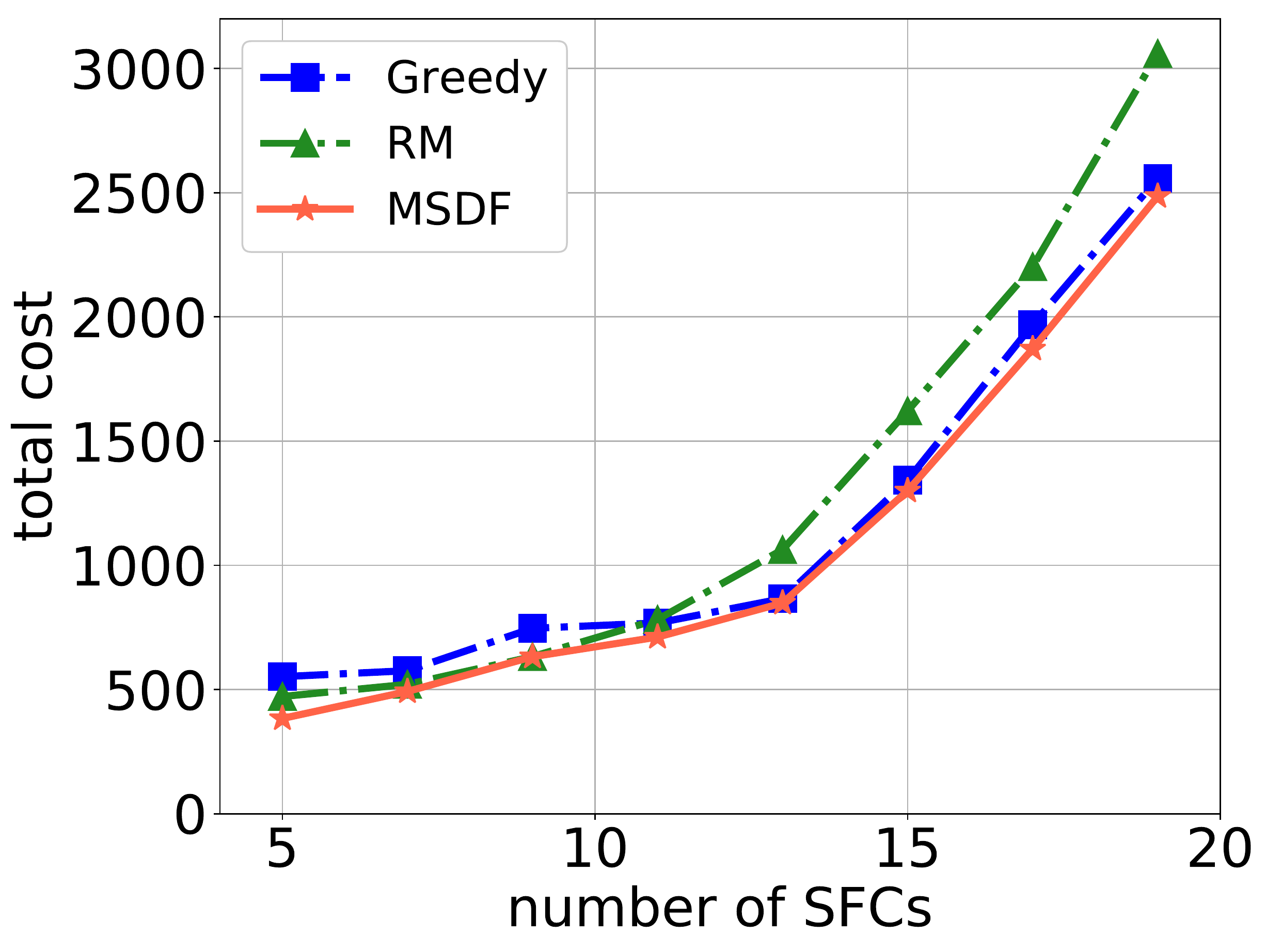}
\label{uu-s-t}
\end{minipage}
}
\subfigure[Number of migrations.]{
\begin{minipage}[t]{0.225\textwidth}
\centering
\includegraphics[width=0.91\textwidth]{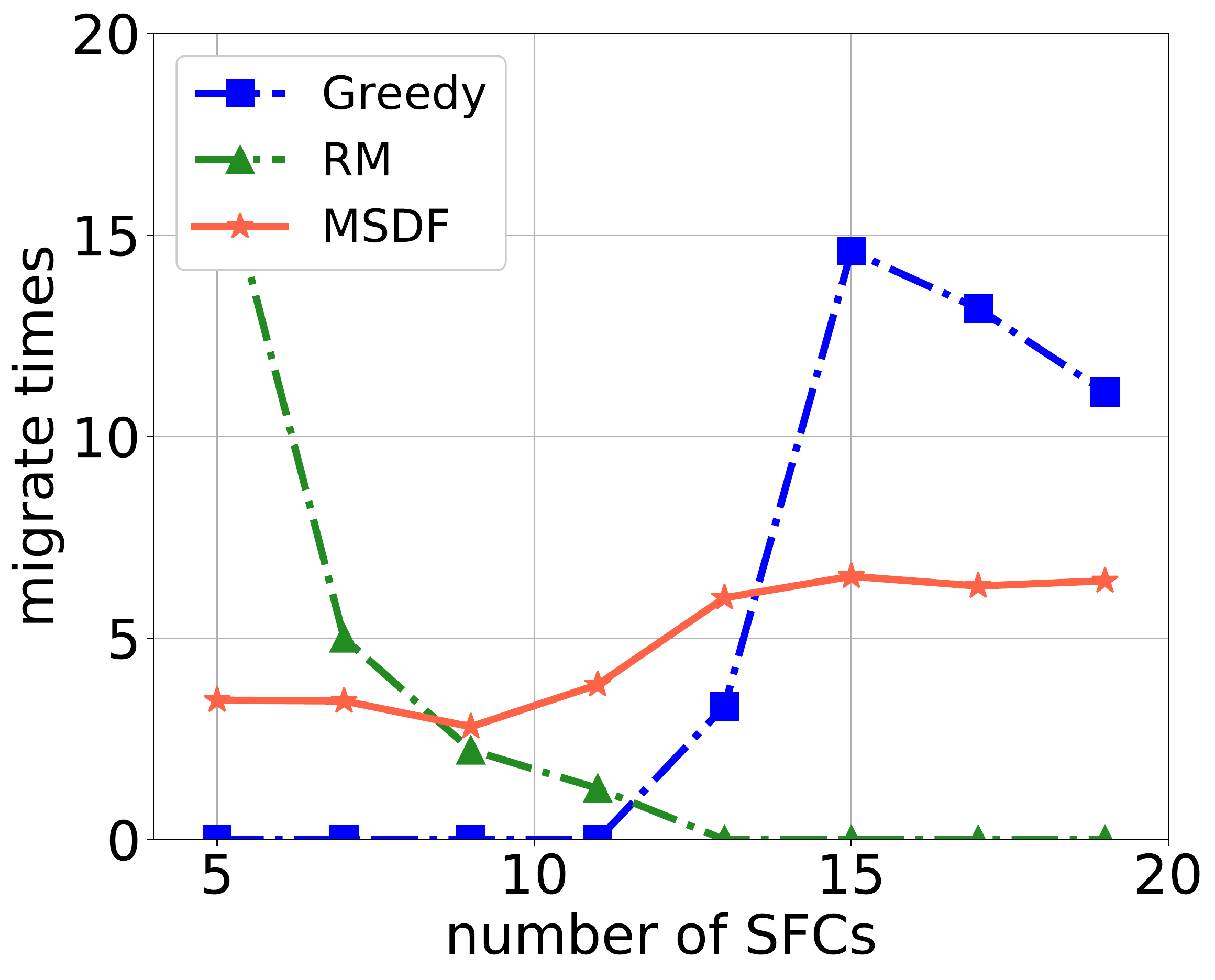}
\label{uu-s-m}
\end{minipage}
}
\subfigure[The degree of overload on nodes.]{
\begin{minipage}[t]{0.225\textwidth}
\centering
\includegraphics[width=0.93\textwidth]{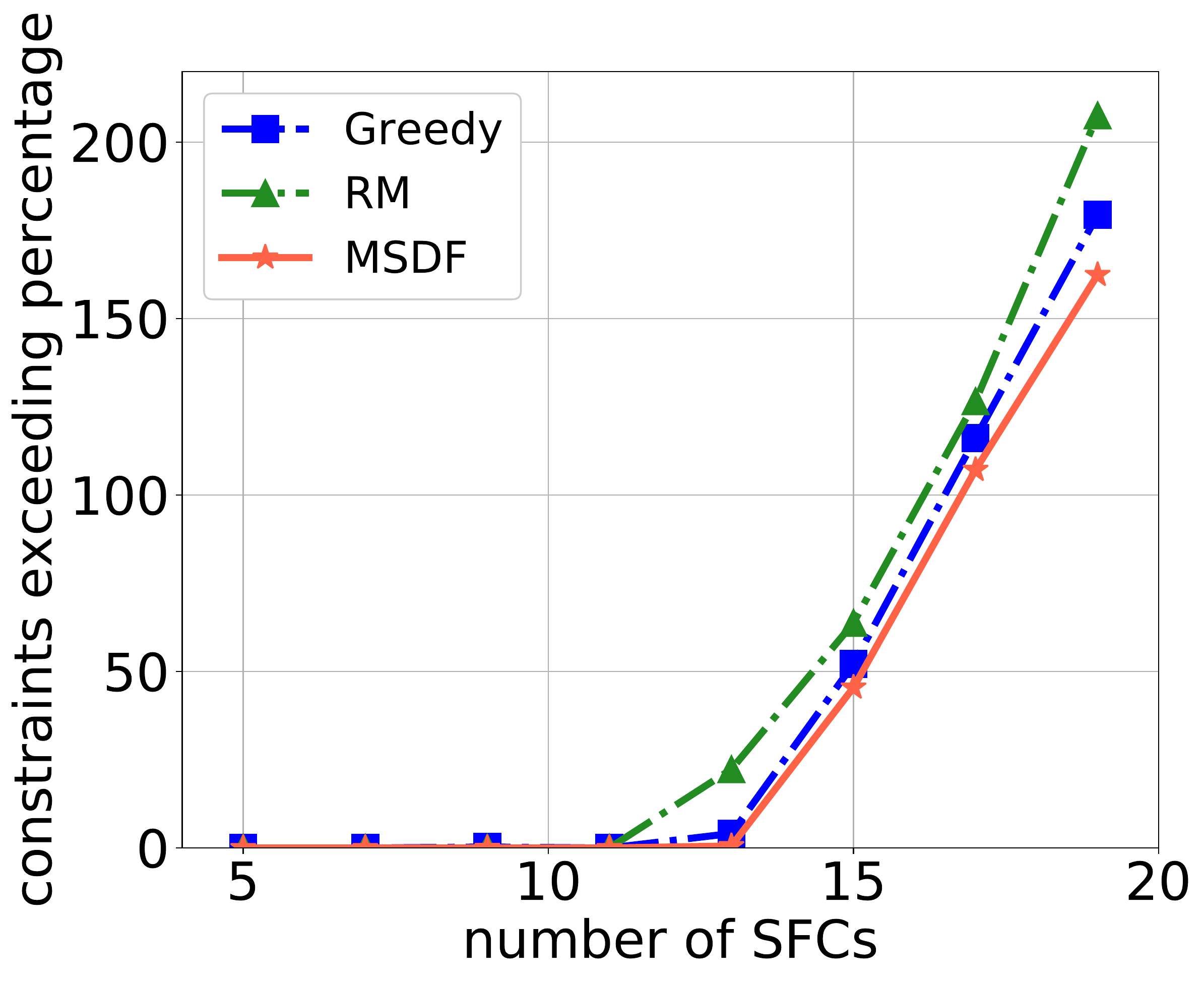}
\label{uu-s-o}
\end{minipage}
}
\subfigure[Basic energy cost.]{
\begin{minipage}[t]{0.225\textwidth}
\centering
\includegraphics[width=0.95\textwidth]{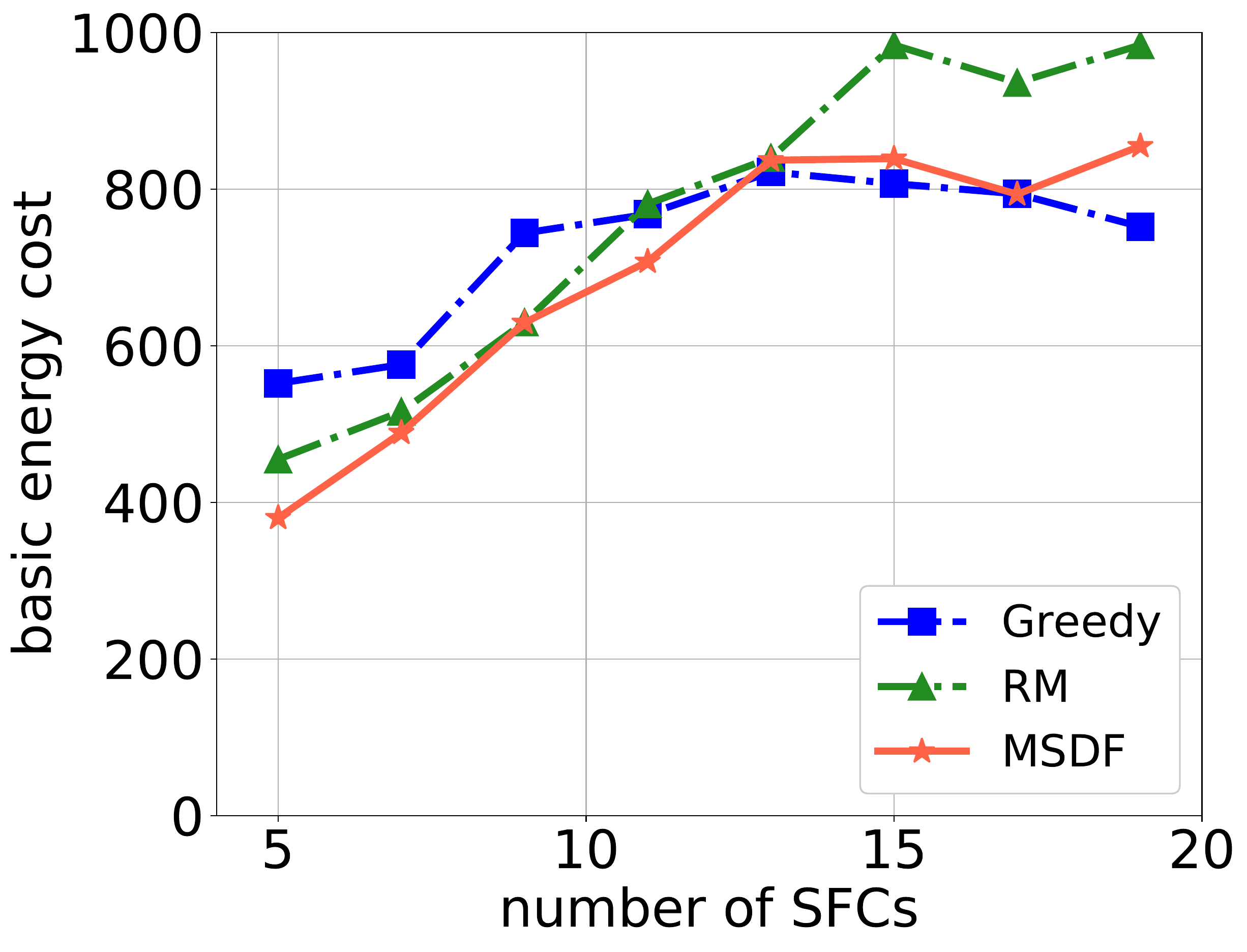}
\label{uu-s-e}
\end{minipage}
}
\centering
\caption{Performance changing along number of SFCs.}
\label{uu-s}
\end{figure*}

Combining migration strategies of subagents without conflicts to approximate the optimal joint migration strategy requires a coordination mechanism. We design a cooperative framework, MSDF (Monitor and Successive Decision Framework), where multi-agents successively make decisions to facilitate minimization of total cost. The basic ideas behind this framework are elaborated as follows:

\emph{1) Cooperative multi-agent framework.} Consider a network with $N$ functional nodes and $S$ SFCs. For the $q_{th}$ SFC with $g_q$ VNFs, we use DQN to choose a VNF and select one node from the other $N-1$ available nodes to place the VNF. In this case, the total number of migration strategies for the entire network is $\prod_{q=1}^S\big(g_q(N\!-\!1)\!+\!1\big)$, which is a huge action space even for a small number of SFCs. Therefore, we decompose the huge action space into smaller ones of subagents designed as Sec. \ref{SSFCM}. The global energy consumption in the reward function (\ref{eqn:6}) directs all subagents to converge toward a common target, while the migration overhead is independent for each subagent. Therefore, collaborative behaviors are generated among these subagents.

\emph{2) Successive decision structure.} To avoid conflict, these subagents make decisions successively. Each subagent makes decision based on the new intermediate network state after the previous subagent executed its action (the state transition is illustrated in Fig. \ref{multi}). Combined with the appropriately designed reward function, the whole network will converge when the last subagent converges. The decision order of subagents depends on the probability of overload, calculated as $P_{q,t}^{SFC}\!=\!\prod_{m\in V_q}\sum_{i\in N^P}y_{i,q,m}^t P_{i,t}^{node}$, where $P_{i,t}^{node}\!\!=\!\!\big(\sum_{v\in F}\!\sum_{q\in S}B_q/Len\sum_{m\in V_q}y_{i,q,m}^th_{v,q,m}t_v^p\big)/C_i$ represents the probability that node $i$ happens to overload. That is to say, the SFC which is most likely to be overloaded will be migrated at first. Finally, the joint migration strategy is formed by concatenating actions of subagents.

\emph{3) Simulation and monitor.} To reduce the overhead caused by frequent interactions between subagents and the environment, the above successive decision process is performed in a simulated environment within SDN control plane. The simulated network is a snapshot of the real network. The final joint strategy will be applied to the real network in the afore-simulated order to acquire the real rewards. At the same time, the SDN controller monitors the network to obtain the signal to start a new learning process. Thus, the subagents are life-long learning agents thus can automatically keep up with new traffic pattern.


The process of the multiple SFC migration using DQN based MSDF is summarized in Algorithm. \ref{decision-dqn}.


\section{Performance evaluation}

The evaluation is based on Uunet which is the real topology of USA Backbone IP network downloaded from topology-zoo with real network traffic data \cite{2}. The Uunet has 49 nodes and 84 links. We set 10 functional nodes depending on the degree of nodes and 4 types of VNFs in the network. We assume that the actually consumed resource of related VMs, the processing time, the configuration time of related VMs and the deployment time of each VNF depend on the type of the VNF. We evaluate our framework from two aspects: convergence and network performance.



\subsection{Convergence performance}
To illustrate that the multi-agent framework can effectively reduce the action space and consequently speed up the convergence, we first compare the convergence performance with one-agent under 3 SFCs of length 3. The size of action space is around 27,000 for the one-agent but 28 for each subagent in MSDF. Their convergence curves are shown in Fig. \ref{conv12m}, from which we can clearly see the difference in terms of convergence speed. After 20,000 rounds of training, the one-agent still does not converge to the same level as MSDF, which only needs hundreds of iterations to converge.

Then we try to find out the main factors affecting the convergence of MSDF. We train the subagents with different parameters, as shown in Fig. \ref{conver}. With other factors fixed, changing the number of SFCs, i.e., the number of cooperative subagents, has slight impact on the convergence performance (Fig. \ref{convsfcnum}). Nevertheless, MSDF has shown its scalability to a larger network with more SFCs as it can still converge fast in the case of 20 SFCs. Fig. \ref{convsfclen} shows that the convergence performance gets deteriorated when the SFCs need to traverse more VNFs, which leads to larger action space. Increasing the number of flows included in each SFC (Fig. \ref{convflownum}) shows graceful convergence performance under different load scenarios, which validates the ability of trading off among different targets.

\subsection{Network cost evaluation}

We compare the converged MSDF with two heuristic algorithms. One is the Greedy proposed in \cite{Optimized}, which trys to ease the overload of functional nodes. Specifically, it migrates VMs in descending order of resource requests on the most overloaded nodes to those nodes that have adequate resource, with satisfying end-to-end delay constraint. The other one is the RM \cite{real}, which performs real-time migration with the target to reduce the end-to-end delay of SFCs under the resource constraint of function nodes.



Firstly, we set 10 SFCs with length of 3, and change the number of flows from 60 to 200 (there are at most 2,000 flows in the network simultaneously). Under the light-load scenario, RM migrates as the resource is sufficient. The result shows that MSDF can save basic energy cost (Fig. \ref{uu-f-e}) by aggregating traffic loads to lesser VMs and nodes with fewer migrations, while the heuristic RM adopts instantaneous optimal solutions leading to frequent migrations (Fig. \ref{uu-f-m}). Along with the increasing loads, RM stops migrating which illustrates that it can not handle the heavy-load situation. Meanwhile, MSDF can hold equivalent level of overload (Fig. \ref{uu-f-o}, where the ordinate is the sum of the scaled percentages) as Greedy with fewer migrations. Moreover, Greedy migrates only under the heavy-load scenario, as shown in (Fig. \ref{uu-f-m}). In a total word, the proposed framework can adjust automatically to adapt to the dynamic network. Thus it can achieve the smallest total cost from a long-term perspective.

Then we fix the number of flows as 50, and change the number of SFCs from 5 to 19, trying to further demonstrate the scalability of MSDF. The results (Fig. \ref{uu-s}) imply that MSDF can still find the balance between basic energy cost and users' QoS under the dynamic network. Particularly when there are $19$ SFCs in the network, these 19 subagents cooperate successfully to reduce the percentage of unsatisfied QoS constraints with a certain number of migrations (Fig. \ref{uu-s-o}, \ref{uu-s-m}). Consequently, MSDF is still effective when it scales to a larger network.



\section{Conclusion}
In this paper, we formulate the multiple SFC migration problem under dynamic traffic with the goal to minimize the total network cost in a long time span. A novel learning framework MSDF is proposed to handle the high complexity when apply RL to make joint SFC migration decisions. MSDF accelerates the convergence speed and facilitates cooperation among subagents to reduce the total network cost. Simulation results validate the effectiveness of MSDF. Its convergence speed is more than 25 times faster than the convergence speed of a single agent. Meanwhile, it outperforms two typical heuristic algorithms in terms of balancing the network operation cost and users' QoS as well as adapting to dynamic traffic loads.

\bibliographystyle{IEEEtran}
\bibliography{ref}

\end{document}